# Missing short-range interactions in the hydrogen bond of compressed ice

Yongli Huang,[1] Zengsheng Ma[1], Xi Zhang,[2,3] Guanghui Zhou,[4] Yichun Zhou,[1] and Chang Q Sun[1,2,*]

[1] Key Laboratory of Low-dimensional Materials and Application Technology, Ministry of Education, and Faculty of Materials, Optoelectronics and Physics, Xiangtan University, Xiangtan, 411105, China

[2] School of Electrical and Electronic Engineering, Nanyang Technological University, Singapore 639798

[3] College of Materials Science and Engineering, China Jiliang University, Hangzhou 310018, China

[4] Department of Physics and Key Laboratory for Low-Dimensional Structures and Quantum Manipulation (Ministry of Education), Hunan Normal University, Changsha, 410081, China

Combining the Lagrangian-Laplace mechanics and the known pressure dependence of the length-stiffness relaxation dynamics, we have determined the critical, yet often-overlooked, short-range interactions in the O:H–O hydrogen bond of compressed ice. This approach has enabled determination of the force constant, cohesive energy, potential energy of the O:H and the H–O segment at each quasi-equilibrium state as well as their pressure dependence. Evidencing the essentiality of the inter-electron-pair Coulomb repulsion and the segmental strength disparity in determining the asymmetric O:H–O relaxation dynamics and the anomalous properties of ice, results confirmed that compression shortens and stiffens the O:H bond and meanwhile lengthens and softens the H–O bond.

Water and ice has attracted much attention because of its anomalous performance relating to issues from galaxy to geology, climate, biology, and to our daily lives [1-7]. As the building unit, the hydrogen bond (O:H–O)[8] relaxes in different manners under the change of environment conditions, which determines the anomalous properties of water and ice. Contributions have been made experimentally [9-11], computationally [12-16], and theoretically [17, 18] to the understanding of water and ice based on the polarizable or non-polarizable models [19, 20], including the TIP$n$P ($n$ varies from 1 to 5) series [19-23]. Using *ab initio* density functional theory and molecular dynamics calculations, one is able to reproduce some of the anomalies demonstrated by compressed ice with limited knowledge about the nature of the inter- and intra-molecular interactions [24].

The objective of this work is to explore analytically the energy relaxation dynamics of the segmented O:H–O bond of ice under compression based on the Lagrangian-Laplace mechanics [25-27]. With the known length-stiffness relaxation dynamics of the O:H–O bond under compression [3-6] as input, we have been able to determine the force constants, the potential well depths, and the cohesive energies of each part of the O:H–O bond as well as their pressure dependence.

A linear hydrogen bond is assumed for simplicity because the O:H–O bond angle in ice is valued at 170 ± 4° [28]. By averaging the surrounding background interactions of $H_2O$ molecules and protons and the nuclear quantum effect on fluctuations [29], we focus on the short-range interactions in this O:H–O bond with H being the coordination origin. As illustrated in Figure 1, the van der Waals force is limited to the O:H bond (denoted L) [30], the exchange interaction is within the H–O polar-covalent bond (denoted H) [31], and the Coulomb repulsion (denoted C) applies between the electron pairs attached to the adjacent oxygen ions, see Supplementary information (SI) [32].

Because of the short-range nature of the interactions, only the solid lines in the shaded area in Figure

1 are effective for the basic O:H–O unit. These interactions will switch off immediately outside the O:H–O region. These interactions determine the physical properties irrespective of the phase structures of the hydrogen-bonded networks but only O—O interaction bridged by H. The presence of inter-electron-pair Coulomb repulsion dislocates both O ions slightly away from their respective equilibrium position. $\Delta_x$ ($x$ = L for the O:H and $x$ = H for the H–O bond) denotes the dislocations. $d_{x0}$ is the interionic distance at equilibrium without the Coulomb repulsion being involved. $d_x = d_{x0} + \Delta_x$ is the quasi-equilibrium bond length with the Coulomb repulsion being involved. The Coulomb repulsion raises the cohesive energies of the O:H and the H–O from $E_{x0}$ to $E_x$ by the same amount.

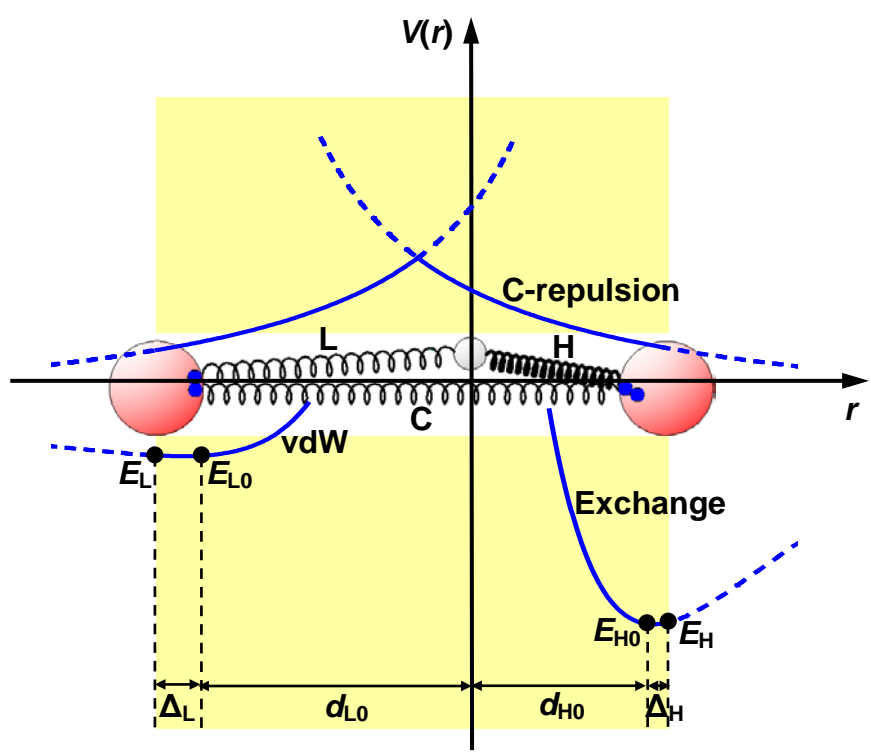

Figure 1 Schematic of the segmented O:H–O bond with springs representing the short-range interactions with H atom being the coordination origin: intramolecular exchange interaction limited to the H–O bond (H), intermolecular van der Waals (vdW) force limited to the O:H bond (L), and the inter-electron-pair Coulomb repulsion (C-repulsion) force between adjacent O—O (C). The red and grey spheres denote the oxygen and the hydrogen atoms, respectively. The pairs of dots on oxygen

denote the electron pairs (lone pair and bonding pair). The Coulomb repulsion pushes both O atoms away from their equilibrium positions.

The O:H-O bond is taken as two oscillators coupled by Coulomb interaction. The reduced mass of the $H_2O$:$H_2O$ oscillator is $m_L = 18 \times 18/(18+18)m_0 = 9m_0$ and that of the H–O oscillator is $m_H = 1 \times 16/(1+16)m_0 = 16/17\ m_0$ with $m_0$ being the unit mass of $1.66 \times 10^{-27}$ kg. The O:H–O bond motion follows the Lagrangian motion equation [25]:

$$\frac{d}{dt}\left(\frac{\partial L}{\partial (dq_i/dt)}\right) - \frac{\partial L}{\partial q_i} = Q_i$$

(1)

The Lagrangian $L = T - V$ consists of the total kinetic energy $T$ and the total potential energy $V$. $Q_i$ denotes the generalized non-conservative forces. Here, it is the pressure $f_P$. The time-dependent $q_i(t)$ represents the generalized variables, denoting the deviating displacements from the equilibrium position of the springs L and H here, i.e. $u_L$ and $u_H$. The kinetic energy $T$ consists of two terms, as the H is taken as the coordination origin,

$$T = \frac{1}{2}\left[m_L\left(\frac{du_L}{dt}\right)^2 + m_H\left(\frac{du_H}{dt}\right)^2\right]$$

(2)

The potential energy $V$ is composed of three terms [32]: the van der Waals interaction $V_L(r_L) = V_L(d_{L0} - u_L)$, the exchange interaction $V_H(r_H) = V_H(d_{H0} + u_H)$, and the Coulomb repulsion $V_C(r_C) = V_C(d_{C0} - u_L + u_H) = V_C(d_C - u_C)$. Here, $d_{C0} = d_{L0} + d_{H0}$ is the distance between the adjacent oxygen ions at equilibrium. $d_C = d_L + d_H$ denotes that distance at quasi-equilibrium. $u_C = u_L + \Delta_L - u_H + \Delta_H$ shows the change of the distance between the two oxygen ions at quasi-equilibrium. The $u_L$ and $u_H$ are assumed to be of the opposite sign because of the O:H and H-O dislocate in the same direction [24]. A harmonic approximation of the potentials by omitting the higher-order terms in their

Taylor's series yields,

$$V = V_L(r_L) + V_H(r_H) + V_C(r_C)$$

$$= \sum_n \left\{ \frac{d^n V_L}{n! dr_L^n}\bigg|_{d_{L0}} (-u_L)^n + \frac{d^n V_H}{n! dr_H^n}\bigg|_{d_{H0}} (u_H)^n + \frac{d^n V_C}{n! dr_C^n}\bigg|_{d_C} (-u_C)^n \right\}$$

$$\approx [V_L(d_{L0}) + V_H(d_{H0}) + V_C(d_C)] - V_C' u_C + \frac{1}{2}[k_L u_L^2 + k_H u_H^2 + k_C u_C^2]$$

(3)

where $V_x(d_{x0})$, commonly denoted $E_{x0}$, is the potential well depths ($n = 0$ terms) of the respective bond. Noting that the Coulomb potential never has an equilibrium point where the repulsion force is 0, we can then expand this potential at quasi-equilibrium point. Therefore, the terms of $n = 1$ is the force equaling 0 for the L and H segments at equilibrium, while equaling $-V_C' u_C$ for the C spring at quasi-equilibrium. Here, $V_C'$ denotes the first order derivative at the quasi-equilibrium position, i.e. $(dV_C/dr_C)|_{d_C}$. The coefficients of the $n = 2$ terms, or the curvatures of the respective potentials, denote the force constants, i.e., $k_x = d^2 V_x / dr_x^2 \big|_{d_{x0}}$ for harmonic oscillators. The $n \geq 3$ terms are the high-order nonlinear contributions that are insignificant, as it will be shown.

Substituting Eqs (2) and (3) into (1) leads to the coupled Lagrangian equation,

$$\begin{cases} m_L \dfrac{d^2 u_L}{dt^2} + (k_L + k_C) u_L - k_C u_H + k_C(\Delta_L - \Delta_H) - V_C' - f_P = 0 \\ m_H \dfrac{d^2 u_H}{dt^2} + (k_H + k_C) u_H - k_C u_L - k_C(\Delta_L - \Delta_H) + V_C' + f_P = 0 \end{cases}$$

(4)

A Laplace transformation [32] turns out solutions to Eq (4),

$$\begin{cases} u_L = \dfrac{A_L}{\gamma_L} \sin \gamma_L t + \dfrac{B_L}{\gamma_H} \sin \gamma_H t \\ u_H = \dfrac{A_H}{\gamma_L} \sin \gamma_L t + \dfrac{B_H}{\gamma_H} \sin \gamma_H t \end{cases}.$$

(5)

The coefficients denote the vibrational amplitudes. $\gamma_L$ and $\gamma_H$ are the vibration angular frequencies of the respective segment, which depend on the force constants and the reduced masses of the oscillators [32]. This set of general solutions indicates that the O:H and the H–O segments share the same form of eigen values of stretching vibration. The force constants $k_x$ and the frequencies $\omega_x$ are correlated as follows,

$$k_{H,L} = 2\pi^2 m_{H,L} c^2 \left(\omega_L^2 + \omega_H^2\right) - k_C \pm \sqrt{\left[2\pi^2 m_{H,L} c^2 \left(\omega_L^2 - \omega_H^2\right)\right]^2 - m_{H,L} k_C^2 / m_{L,H}}$$

(6)

where $c$ is the velocity of light travelling in vacuum. Omitting the Coulomb repulsion, the coupled oscillators will be degenerated into the independent $H_2O:H_2O$ and H–O oscillators with respective vibration frequencies of $\sqrt{k_L / m_L}$ and $\sqrt{k_H / m_H}$. With the measured $\omega_L$ and $\omega_H$, and the known $k_C$, one can obtain the force constants $k_x$, the potential well depths $E_{x0}$, and the cohesive energy $E_x$.

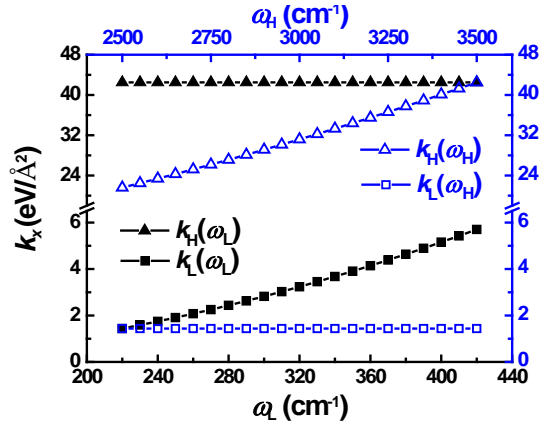

Figure 2 $\omega_L$ and $\omega_H$ dependence of the $k_L$ and $k_H$ with $k_C = 0.17$ eV/Å$^2$. $k_L$ increases from 1.44 to 5.70 eV/Å$^2$ while the $k_H$ increases from 21.60 to 42.51 eV/Å$^2$ as their respective frequency increases. The $k_L(\omega_H)$ and the $k_H(\omega_L)$ remains almost constant.

The force constant due to Coulomb repulsion is, $k_C = q_+ q_- / \left(2\pi \varepsilon_r \varepsilon_0 d_C^3\right)$ at equilibrium. Here, $\varepsilon_r$ is the relative dielectric constant of ice, equaling to 3.2. $\varepsilon_0 = 8.85 \times 10^{-12}$ F/m, is the vacuum dielectric constant. The $q_: = 2e$ for the electron lone pair, and $q_- = 0.2e$ or so, is the effective charge referring to our density functional theory optimizations [32]. In this situation, the $k_C$ equals to 0.17 eV/Å² at 0 GPa. The $\omega_L$ and $\omega_H$ dependence of the $k_L$ and the $k_H$, in Figure 2, shows that the $k_L$ increases from 1.44 to 5.70 eV/Å² while the $k_H$ increases from 21.60 to 42.51 eV/Å² with their respective frequency. The $k_L(\omega_H)$ and the $k_H(\omega_L)$ remains, however, almost constant. Therefore, Eq. (6) can be simplified as,

$$k_{H,L} = 4\pi^2 c^2 m_{H,L} \omega_{H,L}^2 - k_C$$

(7)

With the measured $\omega_L = 237.42$ cm$^{-1}$ and $\omega_H = 3326.14$ cm$^{-1}$ for the ice-VIII phase under the atmospheric pressure [3-6], Eq (7) derives $k_L = 1.70$ eV/Å² and $k_H = 38.22$ eV/Å². With the known $d_L = 0.1768$ nm and $d_H = 0.0975$ nm under Coulomb repulsion[24], we can obtain the free length $d_{L0}$ is 0.1628 nm, and the $d_{H0}$ is 0.0969 nm. Then, with the derived values of $k_L$ and $k_H$, as well as the $E_{H0} = 3.97$ eV [32], we can determine the parameters in the van der Waals and the Morse potentials [32], as well as the force fields of the O:H–O bond at the ambient pressure,

$$\begin{cases} k_L = 72 E_{L0}/d_{L0}^2 = 1.70\,\text{eV}/\text{A}^2 \\ k_H = 2\alpha^2 E_{H0} = 38.22\,\text{eV}/\text{A}^2 \end{cases}$$

or

$$\begin{cases} E_{L0} = 1.70 \times 1.628^2/72 = 0.062\,\text{eV} \\ \alpha = (38.22/3.97/2)^{1/2} = 2.19\,\text{A}^{-1} \end{cases}$$

Using the measured [3-6, 24] Raman shifts $\omega_x$ and the interionic distances $d_x$ [32] as input, we can readily calculate the evolution of the force constant and cohesive energy of the respective segment, from one quasi-equilibrium to another, under compression based on Eq. (6). Table 1 and Figure 3 display the results.

Results indicate that the compression shortens and stiffens the softer O:H bond, meanwhile,

lengthens and softens the H–O bond slightly through the Coulomb repulsion, which results in contraction of the O—O distance towards O:H and H–O length symmetry [3-6, 24, 33, 34]. The $k_C$ (curvature of the Coulomb potential) in Figure 3(a), keeps almost constant under compression because of the low compressibility of O—O distance. The $k_L$ increases more rapidly than $k_H$ reduces because of the coupling of the compression, the repulsion, and the potential disparity of the two segments.

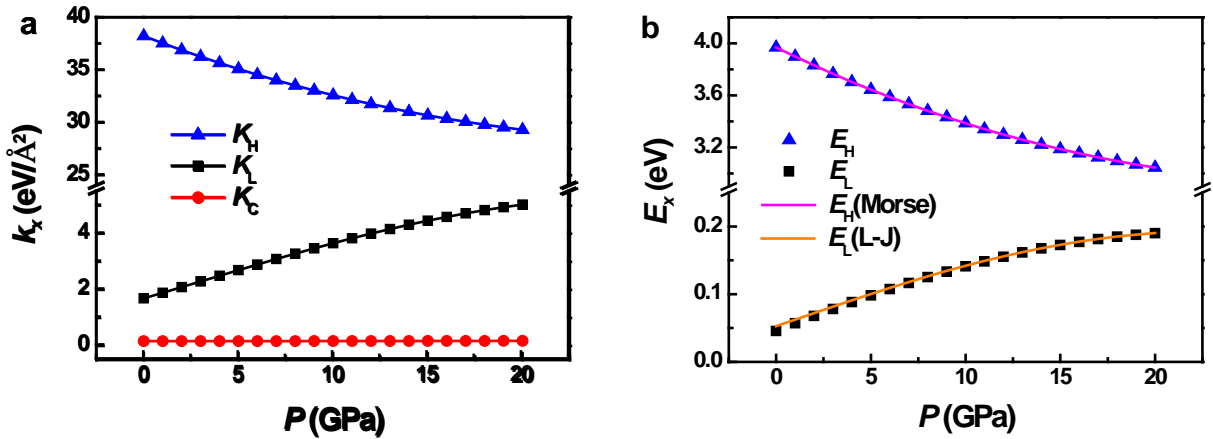

Figure 3 Pressure dependence of (a) the force constant $k_x(P)$ and (b) the cohesive energy $E_x(P)$ of the respective segment of the hydrogen bond. Solid lines in (b) results from the potential functions, which matching well the scattered data of the harmonic approximation.

As the $d_L$ shortens by 4.3% from 0.1768 to 0.1692 nm and the $d_H$ lengthens by 2.8% from 0.0975 to 0.1003 nm with pressure increasing from from 0 to 20 GPa [24]. Figure 3(b) indicates that the increase of pressure from 0 to 20 GPa stiffens the O:H bond from 0.046eV to 0.190 eV while soften the H–O bond from 3.97 eV to 3.04 eV. When the pressure goes up to 60 GPa, the O:H bond almost equals to the elongated H–O bond in length of about 0.110 nm, forming a symmetric O:H–O bond [3-6, 24]. At 60 GPa, the $k_L$ = 10.03 eV/Å$^2$ and $k_H$ = 11.16 eV/Å$^2$, the $E_L$ recovers slightly, see Table 1. Results indicate that the nature of the interaction within the segment remains though the length and force constant approaches to equality, which means that the sp$^3$-hybridized oxygen could hardly be de-hybridized by compression.

Table 1 Pressure dependence of the O:H–O segmental cohesive energy ($E_x$), force constant ($k_x$), and the stepped deviation ($\Delta_x$) from the equilibrium position. Subscript $x$ denotes L and H. The measured $d_x(P)$ and $\omega_x(P)$ [3-6, 24, 32] are used as input in calculations.

| $P$ (GPa) | $E_L$ (eV) | $E_H$ (eV) | $k_L$ (eV/Å$^2$) | $k_H$ (eV/Å$^2$) | $\Delta_L$ (10$^{-2}$ nm) | $\Delta_H$ (10$^{-4}$ nm) |
|---|---|---|---|---|---|---|
| 0  | 0.046 | 3.97 | 1.70  | 38.22 | 1.41 | 6.25 |
| 5  | 0.098 | 3.64 | 2.70  | 35.09 | 0.78 | 6.03 |
| 10 | 0.141 | 3.39 | 3.66  | 32.60 | 0.51 | 5.70 |
| 15 | 0.173 | 3.19 | 4.47  | 30.69 | 0.36 | 5.26 |
| 20 | 0.190 | 3.04 | 5.04  | 29.32 | 0.27 | 4.72 |
| 30 | 0.247 | 2.63 | 7.21  | 25.31 | 0.14 | 3.85 |
| 40 | 0.250 | 2.13 | 8.61  | 20.49 | 0.08 | 3.16 |
| 50 | 0.216 | 1.65 | 9.54  | 15.85 | 0.05 | 2.71 |
| 60 | 0.160 | 1.16 | 10.03 | 11.16 | 0.04 | 3.35 |

Figure 4 shows the $E_x$-$d_x$ asymmetric relaxation dynamics of the O:H–O bond in compressed ice. The oxygen ion (solid spheres in the bottom of Figure 4) in the O:H bond moves towards while the other in the H−O bond away from the H origin. The intrinsic equilibrium position of the oxygen in H−O almost superposes on its quasi-equilibrium position, with the distance of only $6.25 \times 10^{-4}$ nm. However, for O:H, the distance is $1.41 \times 10^{-2}$ nm, evidencing a very soft vdW bond. The cohesive energies of both segments relax along the contours as a resultant of the Coulomb repulsion and the compression.

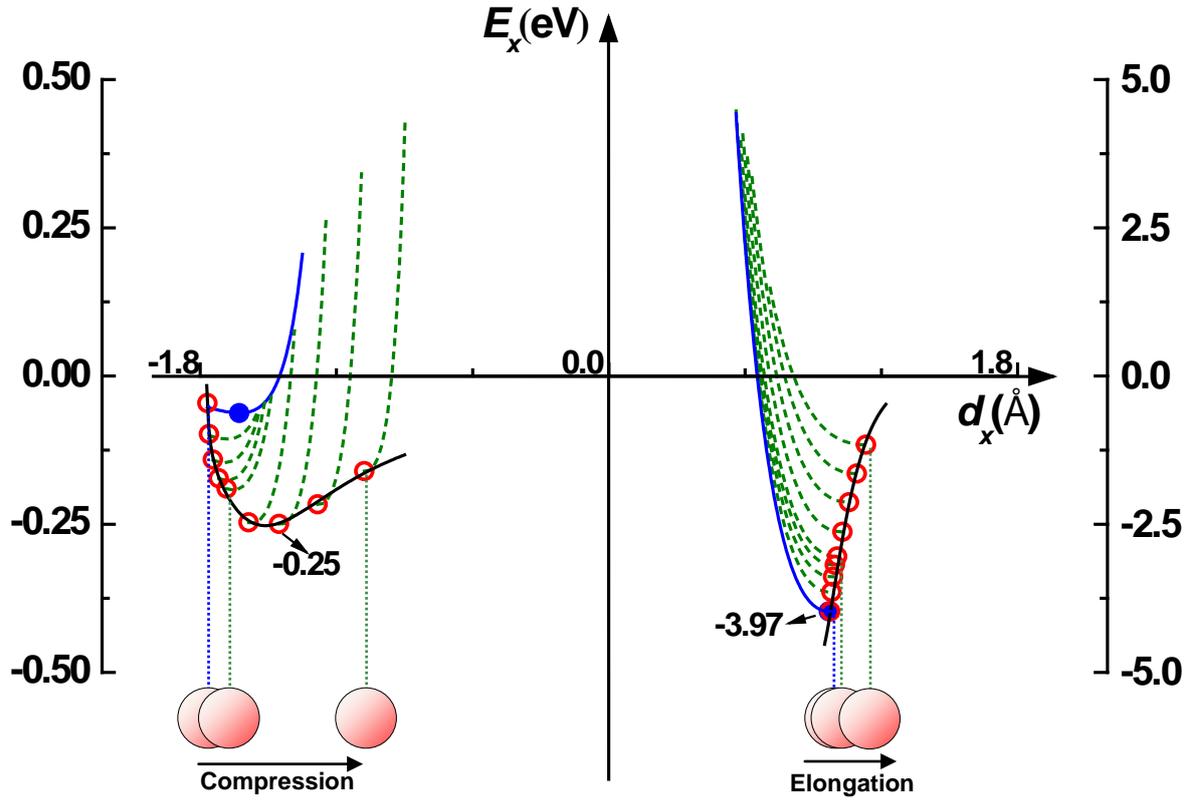

Figure 4  $E_x$-$d_x$ relaxation dynamics of the O:H–O bond of compressed ice (from left to right, $P = 0$, 5, 10, 15, 20, 30, 40, 50, 60 GPa). Small solid circles in blue represent the intrinsic equilibrium coordinates (length and energy) of the oxygen without the Coulomb repulsion, and small open circles denote the quasi-equilibrium coordinates caused by both the Coulomb repulsion and the pressure. The leftmost solid (0 GPa) and the broken curves show the potentials at quasi-equilibrium while the thick solid lines are the contours of the $E_x$-$d_x$ that approach the respective vdW and the Morse potential at equilibrium. Note scale difference between the two segments.

In summary, with the aid of Lagrangian-Laplace mechanics, we have been able to formulate, correlate, clarify, and quantify the short-range interactions in the flexile, polarizable hydrogen bond of compressed ice. This approach has enabled us to determine the cohesive energy, force constant, potential field of each segment and their pressure dependence based on the measurements.

Financial support from National Natural Science Foundation (No. 11172254) of China is acknowledged.

# Supplementary Information

# Missing short-range interactions in the hydrogen bond of compressed ice

Yongli Huang, Xi Zhang, Zengsheng Ma, Guanghui Zhou, Yichun Zhou and CQ Sun

*E-mail:* Ecqsun@ntu.edu.sg

1. **Nomenclature**

| (1) Segmented Hydrogen bond 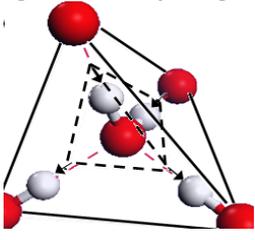 | An expansion of Pauling's Ice Rule results the O:H–O bond that is composed of the intermolecular O:H van der Waals bond and the intramolecular H–O polar-covalent bond, rather than either of them alone [1]. This allows the inclusion of the ultra-short-range interactions discussed here. |
|---|---|
| (2) Electron pair repulsion 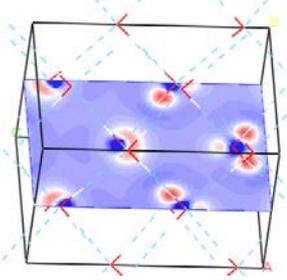 | DFT derived charge distribution in the cross-section of a unit cell of the ice-VIII phase [2]. Red colour represents for electron pairs and the blue for $O^{2-}$ cores. The electron pairs are strongly and eccentrically localized. |

## 2. Potentials for the H-bond ultra-short-range interactions

The short-range interactions include the van der Waals force limited to the O:H bond [3], the exchange interaction in the H–O polar-covalent bond [4], and the Coulomb repulsion between the lone and shared electron pairs attached to the oxygen ions.

$$\begin{cases} V_L(r_L) = V_{L0}\left[\left(\dfrac{d_{L0}}{r_L}\right)^{12} - 2\left(\dfrac{d_{L0}}{r_L}\right)^{6}\right] & \text{(Lennard - Jones potential)} \\ V_H(r_H) = V_{H0}\left[e^{-2\alpha(r_H - d_{H0})} - 2e^{-\alpha(r_H - d_{H0})}\right] & \text{(Morse potential)} \\ V_C(r_C) = \dfrac{q_+ q_-}{4\pi\varepsilon_r\varepsilon_0 r_C} & \text{(Coulomb potential)} \end{cases}$$

(**S1**)

where $V_{L0}$ and $V_{H0}$, commonly denoted $E_{L0}$ and $E_{H0}$, are the potential well depths of the van der Waals bond and the covalent bond, respectively. $r_x$ and $d_{x0}$ ($x$= L, H, and C) denote the interionic distances (corresponding the length of the springs) at arbitrary position and at equilibrium, respectively. $\alpha$ is a parameter controlling the width of the potential well. $\varepsilon_r$ is the relative dielectric constant of ice, equaling to 3.2. $\varepsilon_0 =$ 8.85×10$^{-12}$ F/m, is the vacuum dielectric constant. $q_:$ and $q_.$ denote the charges on respective oxygen ion in two segmented bonds.

With the known Coulomb potential and the measured length-stiffness relaxation parameters [2], the L-J and Morse potentials may be mathematized. Table S1 lists the expressions of the zeroth- to third-derivative of the Taylor series for L-J and Morse potentials. Table S2 gives the corresponding values (energies) of the zeroth- to third-order items evoluting with the pressure. It confirms that the harmonic approximation is suitable because the 3rd item is much smaller than the 2nd item.

Table S1 The zeroth- to third-derivative of the L-J and Morse potentials

| Derivative | L-J potential | Morse potential |
| --- | --- | --- |
| $V_{x0}(E_{x0})$ | $E_{L0}$ | $E_{H0}$ |
| $V_x'$ | 0 | 0 |
| $V_x''$ ($k_x$) | $72E_{L0}/d_{L0}^2$ | $2\alpha^2 E_{H0}$ |
| $V_x'''$ | $-1512E_{L0}/d_{L0}^3$ | $-6\alpha^3 E_{H0}$ |

Table S2 The values for the first four items of the Taylor series of the L-J and the Morse potentials

| $P$ (GPa) | Calculated energy (eV) | | | | | | | |
| --- | --- | --- | --- | --- | --- | --- | --- | --- |
| | L-J potential | | | | Morse potential | | | |
| | 0th | 1st | 2nd (×10$^{-3}$) | 3rd (×10$^{-3}$) | 0th | 1st | 2nd (×10$^{-3}$) | 3rd (×10$^{-3}$) |
| 0 | 0.0625 | 0 | 16.8102 | 10.1750 | 3.9700 | 0 | 0.7465 | 0.0102 |
| 5 | 0.1063 | | 8.2883 | 2.7002 | 3.6447 | | 0.6387 | 0.0085 |
| 10 | 0.1458 | | 4.7185 | 0.9904 | 3.3859 | | 0.5300 | 0.0066 |
| 15 | 0.1755 | | 2.9185 | 0.4391 | 3.1875 | | 0.4247 | 0.0049 |
| 20 | 0.1919 | | 1.9033 | 0.2212 | 3.0450 | | 0.3271 | 0.0034 |
| 30 | 0.2477 | | 0.6599 | 0.0397 | 2.6290 | | 0.1880 | 0.0016 |
| 40 | 0.2498 | | 0.2432 | 0.0089 | 2.1285 | | 0.1022 | 0.0007 |
| 50 | 0.2165 | | 0.0967 | 0.0024 | 1.6465 | | 0.0581 | 0.0003 |
| 60 | 0.1605 | | 0.0697 | 0.0017 | 1.1595 | | 0.0626 | 0.0005 |

## 3. Lagrangian-Laplace solution

With the Lagrangian approximation, the vibration equations for O:H–O hydrogen bond can be deduced as shown in Eq.(4) in the main text.

Letting $(k_L + k_C)/m_L = a$, $(k_H + k_C)/m_H = b$, $k_C/m_L = c$, $k_C/m_H = d$, $[k_C(\Delta_H - \Delta_L) + V_C' + f_P]/m_L = e$, $[k_C(\Delta_H - \Delta_L) + V_C' + f_P]/m_H = f$, this equation becomes,

$$\begin{cases} \dfrac{d^2 u_L}{dt^2} + au_L - cu_H - e = 0 \\ \dfrac{d^2 u_H}{dt^2} + bu_H - du_L + f = 0 \end{cases}$$

(S2)

Assuming that the initial displacements $u_L(0)=u_H(0)=0$, and the initial velocities $du_L/dt\big|_{t=0} = \upsilon_{L0}$, $du_H/dt\big|_{t=0} = \upsilon_{H0}$. Eq. (S2) can be reorganized based on Laplace transformation,

$$\begin{cases} (s^2 + a)U_L - cU_H = \upsilon_{L0} + e/s \\ -dU_L + (s^2 + b)U_H = \upsilon_{H0} - f/s \end{cases}$$

(S3)

where $U_L$ and $U_H$ are the Laplacians of the $u_L$ and $u_H$, respectively, with

$$U_L = U_L(s) = \int_0^\infty u_L(t)e^{-st}dt, \quad U_H = U_H(s) = \int_0^\infty u_H(t)e^{-st}dt$$

where, $s$ is a complex variable. Introducing $\gamma_L = \sqrt{(a+b)/2 - \lambda}$ and $\gamma_H = \sqrt{(a+b)/2 + \lambda}$, where $\lambda = \sqrt{(a-b)^2 + 4cd}/2$, we obtain the solutions to Eq. (S3),

$$\begin{cases} U_L = A_L \dfrac{1}{s^2 + \gamma_L^2} + B_L \dfrac{1}{s^2 + \gamma_H^2} \\ U_H = A_H \dfrac{1}{s^2 + \gamma_L^2} + B_H \dfrac{1}{s^2 + \gamma_H^2} \end{cases}$$

(S4)

where

$$A_L = \frac{c\upsilon_{H0} + b\upsilon_{L0} - \upsilon_{L0}\gamma_L^2}{\gamma_H^2 - \gamma_L^2}; \quad B_L = -\frac{c\upsilon_{H0} + b\upsilon_{L0} - \upsilon_{L0}\gamma_H^2}{\gamma_H^2 - \gamma_L^2};$$

$$A_H = \frac{a\upsilon_{H0} + d\upsilon_{L0} - \upsilon_{H0}\gamma_L^2}{\gamma_H^2 - \gamma_L^2}; \quad B_H = -\frac{a\upsilon_{H0} + d\upsilon_{L0} - \upsilon_{H0}\gamma_H^2}{\gamma_H^2 - \gamma_L^2}.$$

These parameters denote the vibrational amplitudes.

An inverse Laplace transformation of Eq. (S4) results in Eq. (5) in the main manuscript, and the correlation between the frequency and force constants:

$$\begin{cases} \omega_{H,L} = \dfrac{1}{2\pi c}\left\{\dfrac{1}{2m_H m_L}\left[m_H k_L + m_L k_H + (m_L + m_H)k_C \pm \sqrt{[m_L k_H - m_H k_L + (m_L - m_H)k_C]^2 + 4m_L m_H k_C^2}\right]\right\}^{1/2} \\ k_{H,L} = 2\pi^2 m_{H,L} c^2 (\omega_L^2 + \omega_H^2) - k_C \pm \sqrt{[2\pi^2 m_{H,L} c^2 (\omega_L^2 - \omega_H^2)]^2 - m_{H,L} k_C^2/m_{L,H}} \end{cases}$$

and,

$$\begin{cases} \omega_{H,L} = \dfrac{1}{2\pi c}\sqrt{\dfrac{k_{H,L} + k_C}{m_{H,L}}} \\ k_{H,L} = 4\pi^2 c^2 m_{H,L} \omega_{H,L}^2 - k_C \end{cases}$$

by delaminating $\omega_H(k_L)$ and $\omega_L(k_H)$ that make no contribution to the cross terms.

## 4. DFT derived charge sharing in the H–O covalent bond of sized clusters

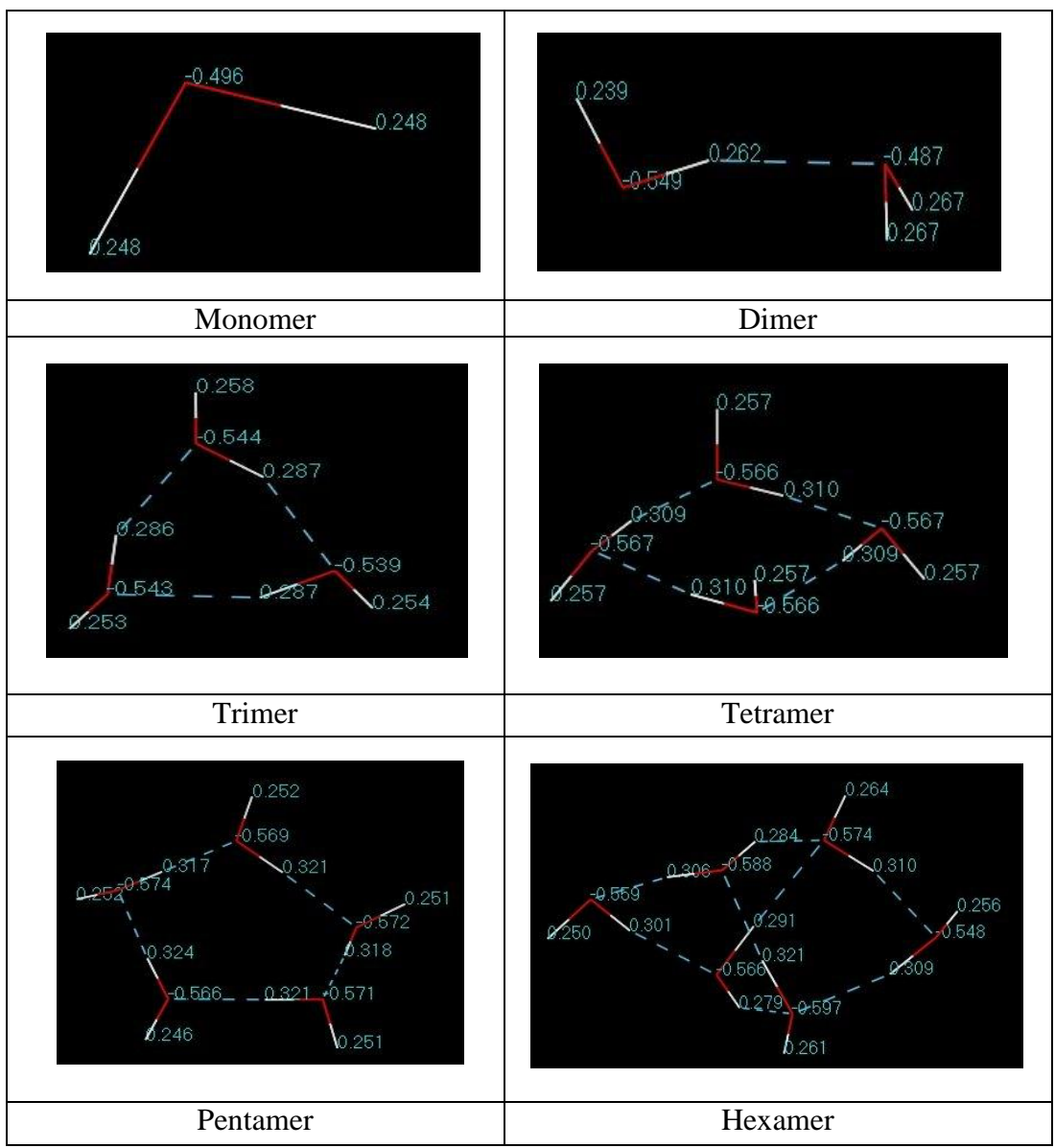

**Fig. S1** DFT derived $(H_2O)_N$ (N=1-6) charge distribution showing the charge taking by the oxygen ion in the covalent bond increases with the cluster size.

## 5. O:H and H–O length relaxation and the *V-P* curve of iceunder compression [2].

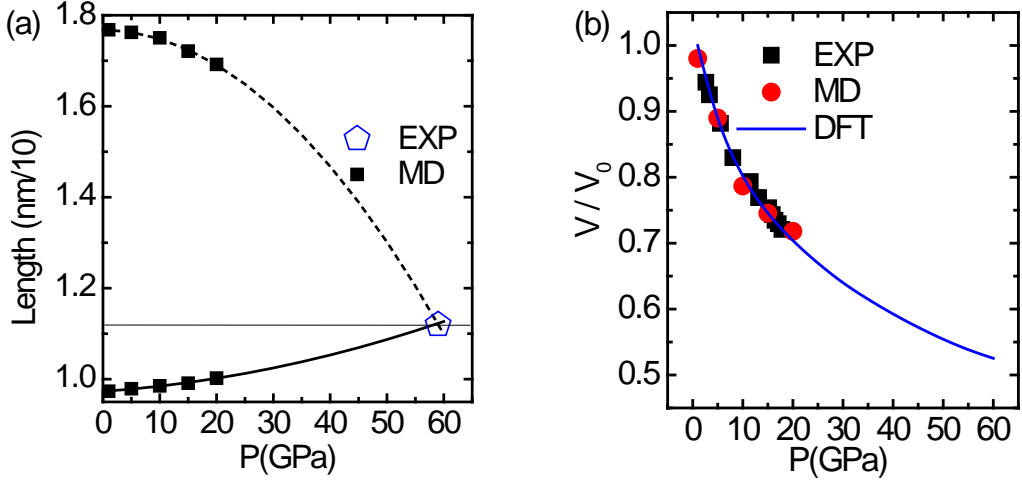

**Fig.S2** (a) MD-derived O:H and H–O asymmetric relaxation dynamics and the proton centralization occurring under 58.6 GPa compression at a O---O distance of 0.221 nm agrees with measurements under 59 GPa at 0.220 nm [12, 13], which decomposes the (b) MD and DFT reproduced [14] *V-P* curve measurements of ice. The curves are formulated by [2]:

$$\begin{pmatrix} d_H/0.9754 \\ d_L/1.7683 \\ V/1.0600 \end{pmatrix} = \begin{pmatrix} 1 & 9.510 & 2.893 \\ 1 & -3.477 & -10.280 \\ 1 & -238.0 & 47.0 \end{pmatrix} \begin{pmatrix} P^0 \\ 10^{-4}P^1 \\ 10^{-5}P^2 \end{pmatrix}$$

## 6. H–O covalent bond energy [2]

From the *P*-dependent critical temperature $T_C$ for the ice VII–VIII phase transition[2], we can obtain the H–O bond energy. The relative shift in $\Delta T_C(P) = T_C(P_0) - T_C(P)$, H–O bond length ($d_H$) and energy ($E_H$) are correlated as follows[2]:

$$\frac{\Delta T_C(P)}{T_C(P_0)} = \frac{\Delta E_x(P)}{E_{x0}} = \frac{-\int_{V_0}^{V} p\, dv}{E_{H0}} = \frac{-s_0 \int_{P_0}^{P} p \frac{dl}{dp} dp}{E_{H0}}$$

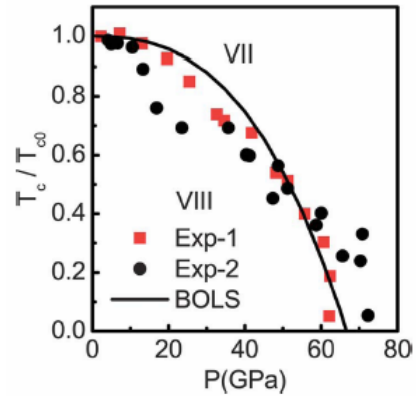

where $l$ can take either form of the $l_L$ or $l_H$ expressed in Fig. S2 caption.

As $P > P_0$, only the segment $dl_H/dp > 0$ satisfies the measured $T_C$-$P$ relation that shows the pressure-depressed $T_C$. Assuming $s_0 = \pi \times (0.053$ nm$)^2$ (the cross-section area of the H–O bond), we obtained $E_{H0} = 3.97$ eV from fitting to the measured pressure-dependent $T_C$ for VII-VIII phase transition[2].

## 7. Pressure-induced Raman shifts [15]

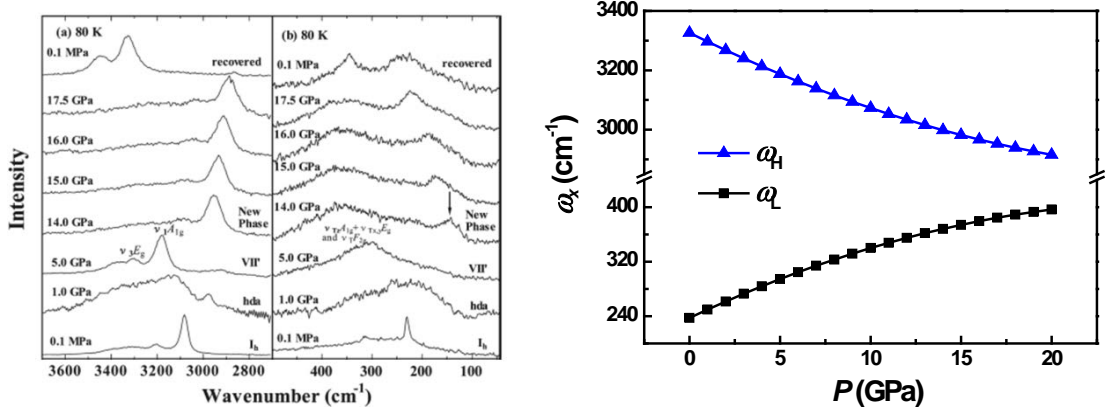

Fig S3 Raman spectra of ice-VIII measured at 80 K as a function of pressure [15], which is formulated as follow,

$$\begin{pmatrix} \omega_H/3326.140 \\ \omega_L/237.422 \end{pmatrix} = \begin{pmatrix} 1 & -0.905 & 1.438 \\ 1 & 5.288 & -9.672 \end{pmatrix} \begin{pmatrix} P^0 \\ 10^{-2}P^1 \\ 10^{-4}P^2 \end{pmatrix}$$

## 8. Pressure-dependent force constants and cohesive energies

The Lagrangian-Laplace derived force constants and cohesive energies can be formulated as:

$$\begin{pmatrix} k_H/38.223 \\ k_L/1.697 \end{pmatrix} = \begin{pmatrix} 1 & -1.784 & 3.113 \\ 1 & 13.045 & -15.258 \end{pmatrix} \begin{pmatrix} P^0 \\ 10^{-2}P^1 \\ 10^{-4}P^2 \end{pmatrix}$$

$$\begin{pmatrix} E_H/3.970 \\ E_L/0.046 \end{pmatrix} = \begin{pmatrix} 1 & -1.784 & 3.124 \\ 1 & 25.789 & -49.206 \end{pmatrix} \begin{pmatrix} P^0 \\ 10^{-2}P^1 \\ 10^{-4}P^2 \end{pmatrix}$$